\documentclass{article}

\usepackage{PRIMEarxiv}

\usepackage[pdftex]{graphicx}
\usepackage[utf8]{inputenc} 
\usepackage[T1]{fontenc}    
\usepackage{hyperref}       
\usepackage{url}            
\usepackage{booktabs}       
\usepackage{amsfonts}       
\usepackage{nicefrac}       
\usepackage{microtype}      
\usepackage{lipsum}
\usepackage{fancyhdr}       
\usepackage{graphicx}       
\usepackage{tabularx}
\usepackage{natbib}
\usepackage{multirow,makecell}
\usepackage{wrapfig}
\usepackage{amssymb}
\usepackage{pifont}
\usepackage{enumitem}
\usepackage{multirow}
\usepackage{booktabs}
\usepackage{tabularx}
\usepackage{xcolor}
\usepackage{listings}

\definecolor{codegreen}{rgb}{0,0.6,0}
\definecolor{codegray}{rgb}{0.5,0.5,0.5}
\definecolor{codepurple}{rgb}{0.58,0,0.82}
\definecolor{backcolour}{rgb}{0.95,0.95,0.92}

\definecolor{codegreen}{rgb}{0,0.6,0}
\definecolor{codegray}{rgb}{0.5,0.5,0.5}
\definecolor{codepurple}{rgb}{0.58,0,0.82}
\definecolor{backcolour}{rgb}{0.95,0.95,0.92}

\lstdefinestyle{mystyle}{
	commentstyle=\color{codegreen},
	keywordstyle=\color{magenta},
	numberstyle=\tiny\color{codegray},
	stringstyle=\color{codepurple},
	basicstyle=\ttfamily\footnotesize,
	breakatwhitespace=false,         
	breaklines=true,                 
	captionpos=b,                    
	keepspaces=true,                 
	numbers=left,                    
	numbersep=5pt,                  
	showspaces=false,                
	showstringspaces=false,
	showtabs=false,                  
	tabsize=2
}

\graphicspath{{media/}}     
\newcommand{\cmark}{\ding{51}}%
\newcommand{\xmark}{\ding{55}}%
\title{CodeTF: One-stop Transformer Library for State-of-the-art Code LLMs}

\author{
	Nghi D. Q. Bui$^*$, Hung Le, Yue Wang, Junnan Li, Akhilesh Deepak Gotmare, Steven C.H. Hoi\thanks{Correspondence: \texttt{\{nghi.bui, shoi\}@salesforce.com}}\\
	Salesforce AI Research \\
\url{https://github.com/salesforce/CodeTF}
}

\begin{document}

\maketitle

\begin{abstract}
Large language models (LLMs) have revolutionized code intelligence, yet practitioners face persistent challenges: fragmented interfaces across models, code-specific preprocessing requirements for diverse programming languages, and a lack of standardized evaluation protocols that hinders reproducibility.

We present CodeTF, an open-source Transformer library for Code LLMs that provides a unified interface supporting 9+ pretrained models across three architectures, code utilities for 15+ programming languages with built-in AST parsing, and standardized evaluation on benchmarks including HumanEval, MBPP, and APPS. The library offers quantization support, parameter-efficient fine-tuning, and comprehensive data preprocessing pipelines.

We describe the design principles, modular architecture, and key components of CodeTF, demonstrating how it simplifies the complete workflow from data preparation to model deployment. CodeTF bridges the gap between machine learning research and software engineering practice, enabling both researchers and practitioners to efficiently develop, evaluate, and deploy Code LLMs.

\end{abstract}


\maketitle

\section{Introduction}
\label{sec:introduction}

\begin{figure}[ht]
	\centering
	\includegraphics[width=1.0\textwidth]{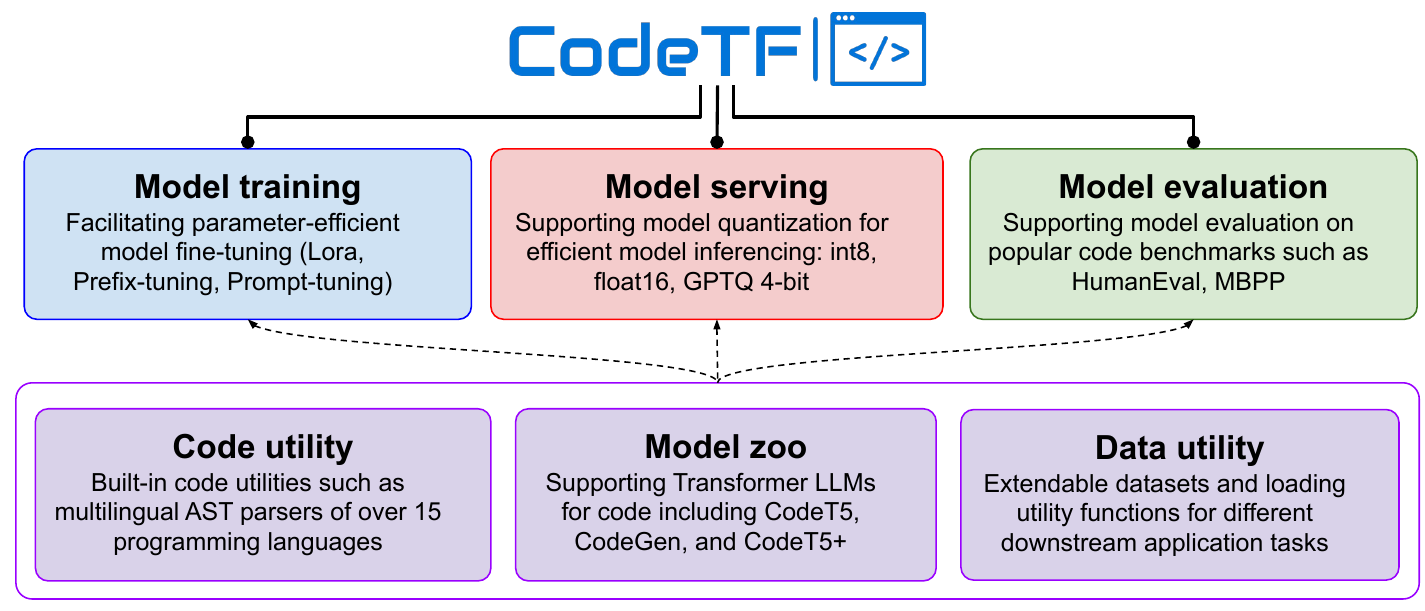}
	\caption{
		\textbf{An overview of CodeTF:}
		A comprehensive Transformer-based library supporting the complete workflow for Code LLMs---from data preparation and model training to serving and evaluation.
		The library integrates pretrained models, code utilities for 15+ programming languages, and standardized benchmarks into a unified framework.
	}
	\label{fig:overview}
\end{figure}

The integration of artificial intelligence into software engineering has accelerated dramatically in recent years, fundamentally transforming how developers write, review, and maintain code. Traditional approaches to code intelligence relied on rule-based static analysis and basic machine learning techniques~\cite{livieri2007very, alexandru2015rapid, chen2017characterizing}. Today, Transformer-based large language models (LLMs) pretrained on massive code corpora---commonly referred to as ``Code LLMs''---have achieved unprecedented performance on tasks ranging from code generation and completion to program repair and code summarization~\cite{codet5, codegen, codebert, graphcodebert, unixcoder, coderl, OpenAI2023GPT4TR}.

The commercial impact of these advances has been substantial. GitHub Copilot, powered by OpenAI's Codex, has been adopted by millions of developers worldwide, demonstrating that Code LLMs can significantly boost developer productivity. Research models such as CodeGen~\cite{codegen}, StarCoder~\cite{starcoder}, and CodeT5+~\cite{wang2023codet5+} have pushed the boundaries of what is possible, achieving impressive results on benchmarks like HumanEval~\cite{codex} and APPS~\cite{apps}. These models learn rich representations of code by training on billions of tokens from open-source repositories on platforms like GitHub~\cite{codegen, codebert, OpenAI2023GPT4TR}, often incorporating programming language features such as data flow and control flow~\cite{codet5, coderl, graphcodebert}.

Despite these advances, practitioners and researchers face significant challenges when working with Code LLMs:

\begin{enumerate}[leftmargin=*]
	\item \textbf{Fragmented Interfaces:} Each model family (CodeBERT, CodeT5, CodeGen, StarCoder) comes with its own API, configuration format, and usage patterns. Switching between models or conducting fair comparisons requires substantial engineering effort to adapt data pipelines and training scripts.

	\item \textbf{Code-Specific Preprocessing:} Unlike natural language, source code has strict syntactic rules that vary across programming languages. Effective preprocessing often requires Abstract Syntax Tree (AST) parsing, identifier extraction, and language-specific tokenization---capabilities not provided by general-purpose NLP libraries.

	\item \textbf{Reproducibility Crisis:} Many published Code LLM results are difficult to reproduce due to inconsistent evaluation protocols, missing preprocessing scripts, and variations in benchmark implementations. This hinders progress and makes it challenging for researchers to build upon prior work.
\end{enumerate}

To address these challenges, we introduce \textbf{CodeTF}, an open-source, comprehensive library that provides a unified framework for developing, training, serving, and evaluating Code LLMs. Figure~\ref{fig:overview} presents an overview of the library architecture. The key insight behind CodeTF is that code intelligence tasks share common infrastructure needs---model loading, quantization, fine-tuning, code parsing, and evaluation---that can be abstracted into reusable, well-designed modules.

CodeTF supports a diverse collection of pretrained models spanning three major architectures: encoder-only models such as CodeBERT~\cite{codebert} and CodeBERTa~\cite{csn}, which are optimized for code understanding tasks; decoder-only models including CodeParrot~\cite{codeparrot}, InCoder~\cite{incoder}, CodeGen~\cite{codegen}, SantaCoder~\cite{santacoder}, and StarCoder~\cite{starcoder}, designed for code generation; and encoder-decoder models like CodeT5~\cite{codet5} and CodeT5+~\cite{wang2023codet5+}, which offer versatility for both understanding and generation.

The library also provides built-in support for popular benchmarks including HumanEval, APPS, MBPP, The Vault~\cite{codex, apps, codexglue, austin2021program, nguyen2023vault}, and CodeXGLUE, with standardized data loaders and evaluation metrics. Through a unified interface, users can load any supported model with a single function call, fine-tune it using parameter-efficient methods (LoRA, Prefix-Tuning, P-Tuning), and evaluate it on multiple benchmarks---all with consistent APIs.

A distinguishing feature of CodeTF is its comprehensive code utility module. Unlike general NLP libraries, CodeTF provides built-in AST parsing for 15+ programming languages using tree-sitter, enabling extraction of code attributes such as function names, identifiers, comments, and control flow structures. These capabilities are essential for tasks like CodeT5's identifier-aware pretraining~\cite{codet5} and are often the most time-consuming aspect of code intelligence research to implement from scratch.

In summary, CodeTF makes four main contributions. First, it provides a \textbf{unified model interface}---a consistent API for loading, serving, and fine-tuning 9+ pretrained Code LLMs across three architectures, with built-in quantization support (8-bit, 4-bit) for efficient deployment. Second, it offers \textbf{comprehensive code utilities} including AST parsing and code attribute extraction for 15+ programming languages, eliminating the need for researchers to build custom preprocessing pipelines. Third, it enables \textbf{standardized evaluation} with built-in support for popular benchmarks (HumanEval, MBPP, APPS, CodeXGLUE) and consistent metrics (pass@k, CodeBLEU, Edit Similarity), facilitating reproducible comparisons. Fourth, it integrates \textbf{parameter-efficient fine-tuning} through HuggingFace PEFT, enabling memory-efficient adaptation of large models using LoRA, Prefix-Tuning, and other techniques. The library is accompanied by extensive documentation, tutorials, and example scripts to help users get started quickly.

CodeTF is designed to serve both researchers exploring new Code LLM architectures and practitioners deploying models in production. By providing a comprehensive, unified, and extensible framework, we aim to accelerate progress in code intelligence research and broaden the adoption of Code LLMs in software development workflows. The following section describes the library's design principles and modular architecture.

\section{Library Design}
\label{sec:principal}

\begin{figure}[t]
	\centering
	\includegraphics[width=1.0\textwidth]{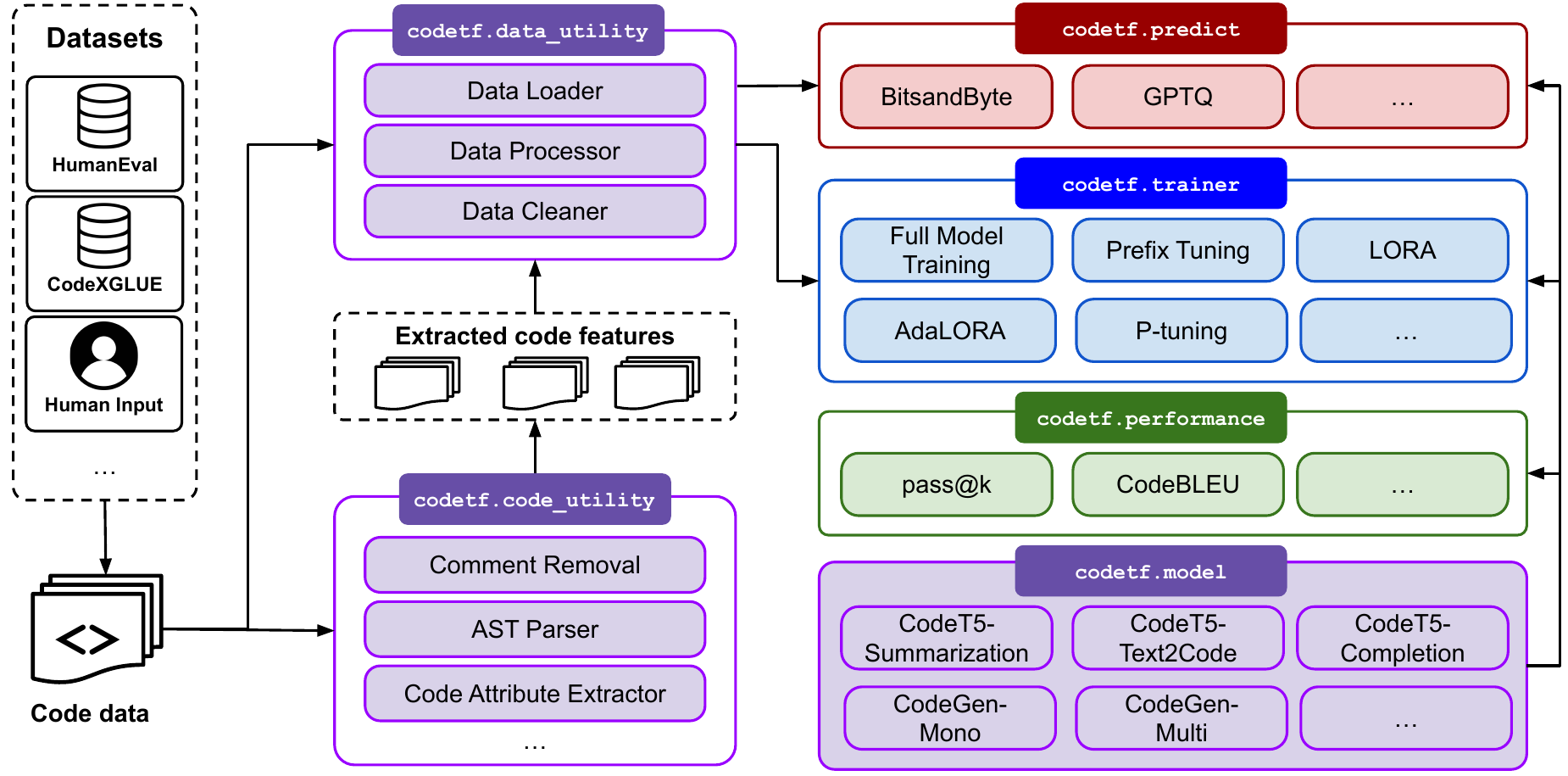}
	\caption{System architecture of CodeTF showing the six core modules: Model Zoo (\texttt{codetf.model}), Model Serving (\texttt{codetf.predict}), Model Training (\texttt{codetf.trainer}), Data Utility (\texttt{codetf.data\_utility}), Code Utility (\texttt{codetf.code\_utility}), and Evaluator (\texttt{codetf.performance}).}
	\label{fig:design}
\end{figure}

Figure~\ref{fig:design} provides a detailed overview of the CodeTF system architecture, illustrating how the modular design enables users to seamlessly combine different components for their specific code intelligence workflows.

\begin{figure}[t]
	\centering
	\includegraphics[width=1.0\textwidth]{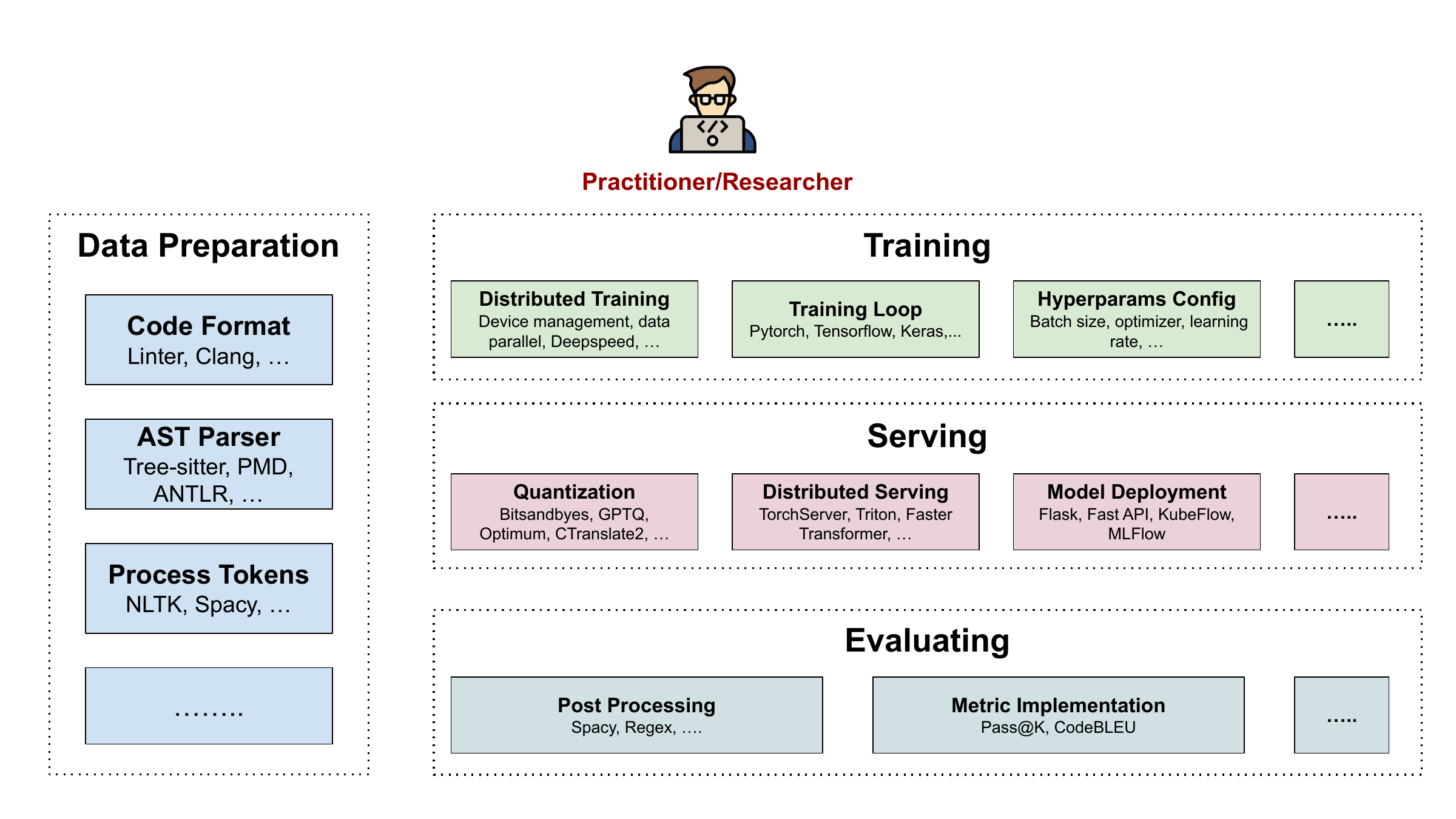}
	\caption{Common workflows when working with Code LLMs. Practitioners typically need to handle four interconnected tasks: data preparation, model training/fine-tuning, model serving, and evaluation. CodeTF provides unified support for all four stages.}
	\label{fig:motivation}
\end{figure}

\subsection{Motivation}

To understand the design rationale behind CodeTF, consider the typical workflow of practitioners working with Code LLMs (Figure~\ref{fig:motivation}). This workflow involves four interconnected tasks, each presenting unique challenges:

\paragraph{Data Preparation.}
Before training or fine-tuning Code LLMs, practitioners must prepare their data through multiple preprocessing steps. This includes formatting source code (using tools like clang-format or ESLint), parsing code into Abstract Syntax Trees (using tree-sitter or language-specific parsers), and extracting relevant features such as function names, identifiers, and comments. These preprocessing steps are crucial---models like CodeT5 require identifier locations for their pretraining objectives, and defect prediction tasks benefit from AST-based features. However, implementing these pipelines from scratch is time-consuming and error-prone.

\paragraph{Training and Fine-Tuning.}
Once data is prepared, users must configure training pipelines including optimizer settings, learning rate schedules, and distributed training across multiple GPUs. For large models (billions of parameters), full fine-tuning may be infeasible, requiring parameter-efficient methods like LoRA or Prefix-Tuning. Each model family often requires different training configurations, making it difficult to experiment across architectures.

\paragraph{Model Serving.}
Deploying trained models for inference presents additional challenges. Large models require quantization (8-bit or 4-bit) to fit in memory and achieve acceptable latency. Users must also handle batching, input preprocessing, and output post-processing consistently across different model architectures.

\paragraph{Evaluation.}
Finally, evaluating models on standard benchmarks requires implementing task-specific metrics (pass@k for code generation, CodeBLEU for translation) and handling benchmark-specific data formats. Inconsistent implementations across papers have led to reproducibility issues in the field.

\vspace{0.5em}
\noindent Existing libraries like HuggingFace Transformers provide excellent general-purpose infrastructure but lack code-specific features such as AST parsing, code-aware tokenization, and specialized evaluation metrics. CodeTF addresses this gap by providing a higher-level abstraction layer specifically designed for code intelligence tasks.

\subsection{Design Principles}

CodeTF is built around six core design principles that guide its architecture and API design:

\begin{enumerate}[leftmargin=*]
	\item \textbf{Comprehensiveness.} CodeTF aims to be a one-stop solution covering the complete Code LLM workflow. The library supports model loading and serving across three architectures (encoder-only, decoder-only, encoder-decoder), multiple fine-tuning strategies (full fine-tuning, LoRA, Prefix-Tuning), comprehensive evaluation metrics, and code utilities for 15+ programming languages. Users should rarely need to integrate external tools for standard code intelligence tasks.

	\item \textbf{Unified Interface.} Despite the diversity of supported models and tasks, CodeTF provides consistent APIs across all components. Loading any model requires a single function call (\texttt{load\_model\_pipeline}) with the same signature regardless of architecture. Training, evaluation, and code parsing follow similar patterns, reducing the learning curve and enabling rapid experimentation across different models.

	\item \textbf{Ease of Use.} CodeTF prioritizes developer experience with minimal setup requirements. Installation via pip or conda provides all dependencies, and users can start with pretrained models in just a few lines of code. Sensible defaults are provided for all configurations, while advanced users can customize any aspect of the pipeline.

	\item \textbf{Extensibility.} The code intelligence field evolves rapidly, with new models and benchmarks emerging frequently. CodeTF follows object-oriented design principles, with abstract base classes (\texttt{BaseTrainer}, \texttt{BaseCodeUtility}) that can be extended for new model architectures or programming languages. Adding support for a new model requires implementing a small number of well-defined interfaces.

	\item \textbf{Scalability.} Training and serving large Code LLMs requires efficient resource utilization. CodeTF integrates with distributed training frameworks (DeepSpeed, FSDP) and provides built-in quantization support (BitsAndBytes, GPTQ) to enable deployment on consumer hardware. The library handles multi-GPU training configuration automatically.

	\item \textbf{Reproducibility.} To address the reproducibility crisis in Code LLM research, CodeTF provides standardized implementations of popular benchmarks and metrics. All evaluation code is versioned and tested, ensuring that results obtained with CodeTF can be reliably reproduced. We also provide reference results for supported models on standard benchmarks.
\end{enumerate}

\subsection{Module Overview}

Based on these principles, CodeTF is organized into six interconnected modules. The \textbf{Model Zoo} (\texttt{codetf.model}) serves as a repository of pretrained and fine-tuned model configurations, supporting encoder-only models (CodeBERT, CodeBERTa), decoder-only models (CodeGen, StarCoder, SantaCoder, InCoder, CodeParrot), and encoder-decoder models (CodeT5, CodeT5+). The \textbf{Model Serving} module (\texttt{codetf.predict}) provides a unified inference interface with built-in quantization (8-bit, 4-bit), batching, and preprocessing for tasks including code generation, completion, and summarization. The \textbf{Model Training} module (\texttt{codetf.trainer}) offers training utilities supporting both full fine-tuning and parameter-efficient methods (LoRA, Prefix-Tuning, P-Tuning, AdaLoRA) with automatic GPU management and checkpoint saving. For data handling and evaluation, the \textbf{Data Utility} module (\texttt{codetf.data\_utility}) provides data loaders for popular benchmarks (HumanEval, APPS, MBPP, CodeXGLUE, The Vault) along with preprocessing pipelines for tokenization and formatting. The \textbf{Code Utility} module (\texttt{codetf.code\_utility}) enables AST parsing for 15+ programming languages using tree-sitter, with utilities for extracting function names, identifiers, comments, and other code attributes. Finally, the \textbf{Evaluator} module (\texttt{codetf.performance}) implements standardized evaluation metrics including pass@k, CodeBLEU, Edit Similarity, and BLEU, with benchmark-specific evaluation scripts. The following section provides detailed descriptions of each module's capabilities and implementation.

\section{Modules and Utilities}
\label{sec:module}

This section provides detailed descriptions of each module in CodeTF, explaining their capabilities, design choices, and usage patterns.

\subsection{Model Zoo}

The Model Zoo (\texttt{codetf.model}) serves as the central repository for model configurations in CodeTF. It provides standardized access to pretrained and fine-tuned checkpoints from state-of-the-art Code LLMs, abstracting away the complexity of different model architectures and hosting platforms.

CodeTF supports three major families of Transformer architectures:
\begin{itemize}[leftmargin=*]
	\item \textbf{Encoder-only models} (CodeBERT~\cite{codebert}, CodeBERTa~\cite{csn}): These models excel at code understanding tasks such as code search, clone detection, and defect prediction. They encode bidirectional context, making them suitable for classification and retrieval tasks.

	\item \textbf{Decoder-only models} (CodeParrot~\cite{codeparrot}, InCoder~\cite{incoder}, CodeGen~\cite{codegen}, SantaCoder~\cite{santacoder}, StarCoder~\cite{starcoder}): Optimized for code generation through autoregressive next-token prediction. These models range from 350M to 16B parameters, offering trade-offs between capability and resource requirements.

	\item \textbf{Encoder-decoder models} (CodeT5~\cite{codet5}, CodeT5+~\cite{wang2023codet5+}): Versatile architectures suitable for both understanding and generation tasks, including code summarization, translation, and refinement.
\end{itemize}

Each model is accompanied by a YAML configuration file specifying the HuggingFace repository URL, tokenizer settings, maximum sequence length, and task-specific parameters. This design enables CodeTF to stay current with the latest model releases simply by adding new configuration files, without requiring code changes.

\begin{figure*}[t]
	\centering
	\includegraphics[width=1.0\textwidth]{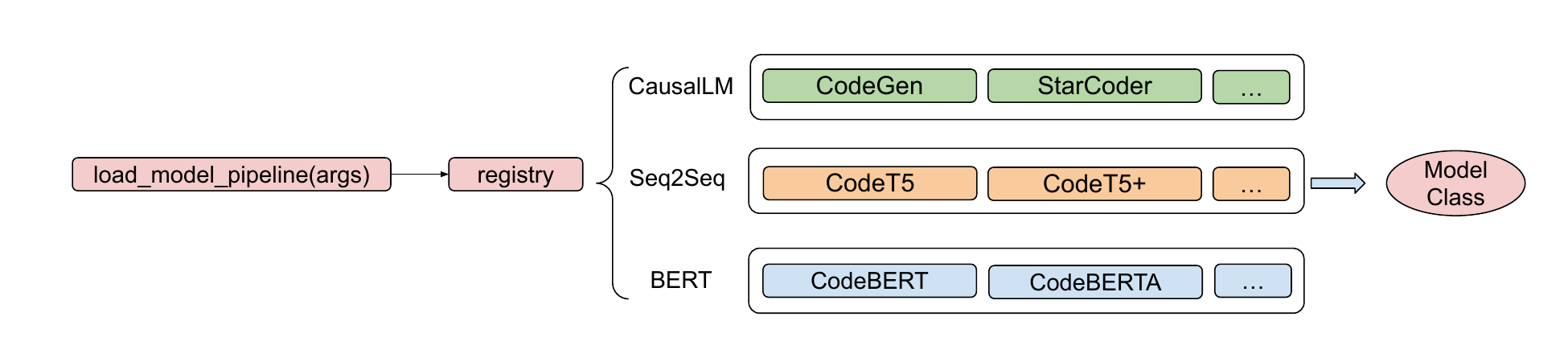}
	\caption{Model loading pipeline in CodeTF. The \texttt{load\_model\_pipeline} function provides a unified entry point that dispatches to architecture-specific model classes (\texttt{CausalLMModel}, \texttt{Seq2SeqModel}, \texttt{BERTModel}) based on the requested model type.}
	\label{fig:registry}
\end{figure*}

\subsection{Model Serving Module}

The Model Serving module (\texttt{codetf.predict}) provides a unified interface for loading models and performing inference. As illustrated in Figure~\ref{fig:registry}, the module uses a registry pattern to dispatch model loading requests to the appropriate model class based on the specified architecture.

\paragraph{Unified Loading Interface.}
Users can load any supported model with a single function call:
\begin{verbatim}
model = load_model_pipeline(
    model_name="codet5-base",
    task="code_summarization",
    language="python"
)
\end{verbatim}
The function automatically handles tokenizer initialization, model weight loading, and device placement (CPU/GPU).

\paragraph{Quantization Support.}
Large models like CodeGen-16B or StarCoder-15B require significant memory and can be slow for inference. CodeTF integrates two quantization backends to address this: BitsAndBytes~\cite{dettmers2022llm}, which enables 8-bit and 4-bit quantization with minimal accuracy loss, reducing memory requirements by 2-4x; and GPTQ~\cite{frantar2022gptq}, which provides post-training quantization for even greater compression, particularly useful for deployment on consumer GPUs. In our benchmarks, 8-bit quantization reduces inference time for CodeGen-16B from approximately 1.2 seconds to 0.4 seconds per sample while maintaining 98\% of the original accuracy.

\subsection{Model Training Module}

\begin{figure}[ht]
	\centering
	\includegraphics[width=1.0\textwidth]{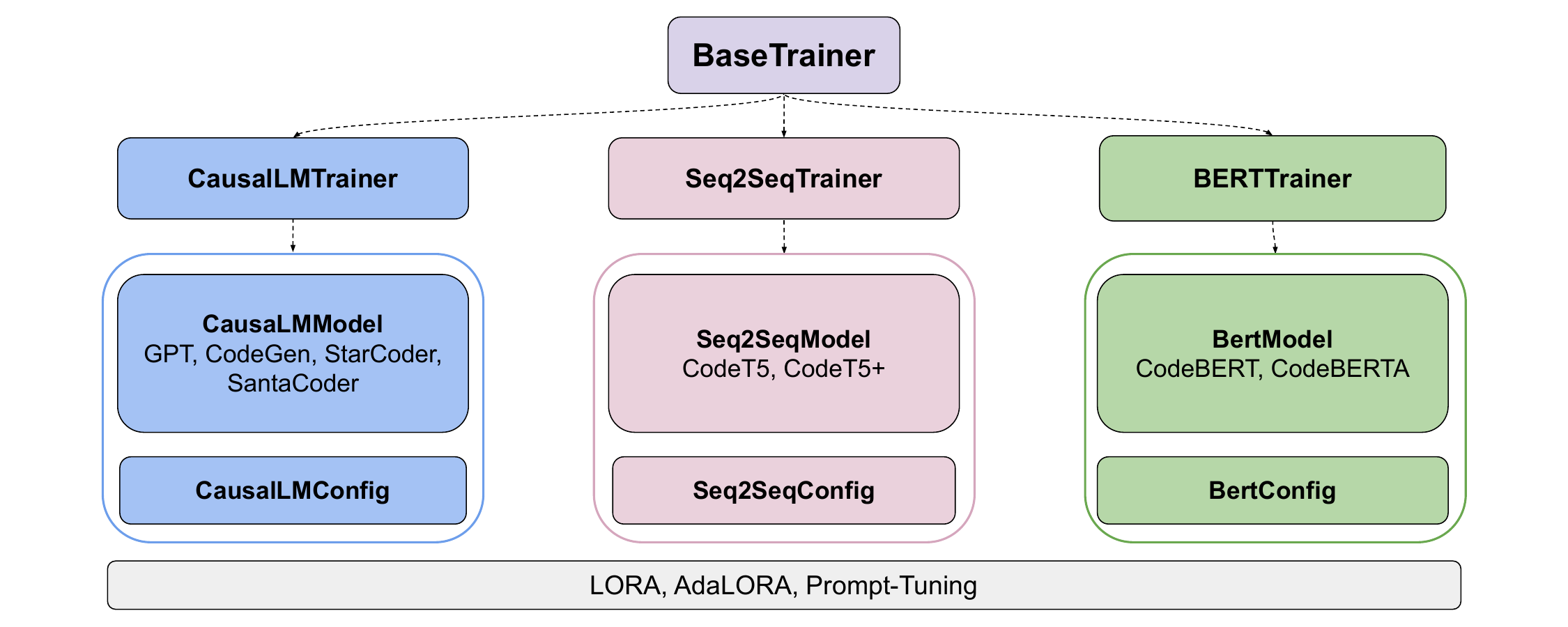}
	\caption{Training module architecture. The \texttt{BaseTrainer} class provides common functionality (checkpoint saving, logging, GPU management) inherited by architecture-specific trainers (\texttt{CausalLMTrainer}, \texttt{Seq2SeqTrainer}, \texttt{BERTTrainer}).}
	\label{fig:trainer}
\end{figure}

The Training Module (\texttt{codetf.trainer}) provides utilities for fine-tuning pretrained models on custom datasets. As shown in Figure~\ref{fig:trainer}, the module follows an inheritance hierarchy with \texttt{BaseTrainer} providing common functionality and specialized trainers handling architecture-specific details.

\paragraph{Full Fine-Tuning.}
For users with sufficient computational resources, CodeTF supports standard full fine-tuning with configurable learning rates, batch sizes, and optimization strategies. The trainer automatically handles multi-GPU training with data parallelism, gradient accumulation for large effective batch sizes, learning rate scheduling (linear, cosine, polynomial decay), checkpoint saving and resumption, and integration with Weights \& Biases for experiment tracking.

\paragraph{Parameter-Efficient Fine-Tuning.}
For large models or limited GPU memory, CodeTF integrates HuggingFace's PEFT library~\footnote{https://github.com/huggingface/peft} to enable parameter-efficient fine-tuning methods:
\begin{itemize}[leftmargin=*]
	\item \textbf{LoRA}~\cite{hu2021lora}: Adds low-rank adaptation matrices to attention layers, typically training only 0.1-1\% of parameters.
	\item \textbf{Prefix-Tuning}~\cite{li2021prefix}: Prepends trainable prefix tokens to each layer's key and value matrices.
	\item \textbf{P-Tuning}~\cite{liu2021gpt}: Uses trainable continuous prompts for task adaptation.
	\item \textbf{AdaLoRA}~\cite{zhang2023adaptive}: Dynamically allocates parameter budget across layers based on importance.
\end{itemize}
These methods enable fine-tuning models with billions of parameters on a single consumer GPU while achieving competitive performance with full fine-tuning.

\subsection{Data Utility Module}

The Data Utility module (\texttt{codetf.data\_utility}) provides comprehensive data loading and preprocessing capabilities for code intelligence tasks.

\paragraph{Benchmark Data Loaders.}
CodeTF includes built-in data loaders for popular code intelligence benchmarks. These include HumanEval~\cite{codex}, which contains 164 hand-written Python programming problems for evaluating code generation; MBPP~\cite{austin2021program}, with 974 Python programming problems covering basic programming concepts; APPS~\cite{apps}, offering 10,000 coding problems of varying difficulty levels with test cases; CodeXGLUE~\cite{codexglue}, a benchmark suite covering 10 code intelligence tasks across multiple languages; and The Vault~\cite{nguyen2023vault}, a comprehensive multilingual dataset for code understanding and generation. Each data loader provides consistent interfaces for accessing training, validation, and test splits, with automatic downloading and caching of datasets.

\paragraph{Preprocessing Pipelines.}
The module offers preprocessing utilities for common code intelligence tasks, including tokenization that wraps HuggingFace tokenizers with code-aware preprocessing (handling special tokens and truncation strategies), code formatting to normalize style and reduce spurious variations in training data, data augmentation with transformations like variable renaming and dead code insertion for robustness training, and batch collation that handles dynamic batching with efficient padding for variable-length code sequences.

\subsection{Evaluator Module}

The Evaluator Module (\texttt{codetf.performance}) provides standardized implementations of evaluation metrics commonly used in code intelligence research.

\paragraph{Execution-Based Metrics.}
For code generation tasks, CodeTF implements the pass@k metric~\cite{codex, alphacode}, which measures the probability that at least one of k generated samples passes all test cases. The evaluator handles secure code execution in sandboxed environments, timeout handling for infinite loops, multi-language support (Python, Java, JavaScript, C++), and parallel evaluation for efficiency.

\paragraph{Reference-Based Metrics.}
For tasks with reference outputs (summarization, translation), CodeTF provides CodeBLEU~\cite{codebleu}, a variant of BLEU that incorporates code-specific features including AST matching and data flow analysis; Edit Similarity~\cite{repocoder}, which measures the Levenshtein distance between generated and reference code; and standard NLP metrics such as BLEU and ROUGE for measuring n-gram overlap. All metrics are implemented with consistent interfaces, enabling easy comparison across different models and tasks.

\subsection{Code Utility Module}

The Code Utility module (\texttt{codetf.code\_utility}) provides language-aware tools for parsing and manipulating source code---a capability that distinguishes CodeTF from general-purpose NLP libraries.

\paragraph{AST Parsing.}
CodeTF integrates tree-sitter\footnote{https://github.com/tree-sitter/tree-sitter}, an incremental parsing library, to provide AST parsing for 15+ programming languages: Java, Python, JavaScript, C, C++, C\#, Go, Ruby, Rust, PHP, Kotlin, Scala, Haskell, Apex, SOQL, SOSL, Solidity, and YAML. Tree-sitter grammars are precompiled into platform-specific binaries (`.so` files) for Darwin, Linux, and Windows, eliminating the need for users to compile parsers themselves.

\paragraph{Code Attribute Extraction.}
Building on the AST infrastructure, the module provides utilities for extracting common code attributes, including identifiers (variable names, function names, class names), comments (docstrings, inline comments, block comments), structural elements (function boundaries, class hierarchies, import statements), and control flow constructs (loop structures, conditional branches, exception handling). These utilities are essential for preprocessing pipelines---for example, CodeT5's pretraining requires masking identifier tokens, which requires accurate identifier extraction.

\paragraph{Language-Specific Utilities.}
Each supported language has a dedicated utility class (e.g., \texttt{PythonCodeUtility}, \texttt{JavaCodeUtility}) that inherits from \texttt{BaseCodeUtility}. This design allows language-specific customizations while maintaining a consistent interface across languages.

The following section demonstrates typical usage patterns for these modules through concrete code examples.

\section{Example Usage}
\label{sec:example}

This section demonstrates CodeTF's capabilities through practical code examples covering the three most common use cases: model inference, fine-tuning, and evaluation.

\paragraph{Model Loading and Inference.}
The following example shows how to load a pretrained CodeT5 model for code summarization and generate a summary for a Python function. With just a few lines of code, users can load a model and perform inference:

\begin{lstlisting}[language=Python]
from codetf.models import load_model_pipeline

summarization_model = load_model_pipeline(model_name="codet5", task="sum-python",
            model_type="base", is_eval=True,
            load_in_8bit=True, weight_sharding=False)

code_snippets = """
    void bubbleSort(int arr[])
    {
        int n = arr.length;
        for (int i = 0; i < n - 1; i++)
            for (int j = 0; j < n - i - 1; j++)
                if (arr[j] > arr[j + 1]) {
                    // swap arr[j+1] and arr[j]
                    int temp = arr[j];
                    arr[j] = arr[j + 1];
                    arr[j + 1] = temp;
                }
    }
"""

# Bubble sort program to sort an integer array
summaries = summarization_model.predict([code_snippets])
\end{lstlisting}

\noindent The \texttt{load\_model\_pipeline} function automatically handles model downloading, tokenizer initialization, and GPU placement. The returned model object provides a simple \texttt{predict} method that accepts raw code as input and returns the generated summary.

\paragraph{Fine-Tuning with Parameter-Efficient Methods.}
CodeTF simplifies the fine-tuning process by providing pre-configured trainers for each model architecture. The following example demonstrates fine-tuning a CodeGen model on the CodeXGLUE code refinement dataset using LoRA for memory efficiency:

\begin{lstlisting}[language=Python]
from codetf.trainer.causal_lm_trainer import CausalLMTrainer
from codetf.data_utility.codexglue_dataset import CodeXGLUEDataset
from codetf.models import load_model_pipeline
from codetf.performance.evaluate import EvaluationMetric

model_class = load_model_pipeline(model_name="causal-lm", task="pretrained",
                model_type="codegen-350M-mono", is_eval=False,
                load_in_8bit=False, weight_sharding=False)


dataloader = CodeXGLUEDataset(tokenizer=model_class.get_tokenizer())
train_dataset, test_dataset, val_dataset = dataloader.load(subset="text-to-code")

evaluator = EvaluationMetric(metric="bleu", tokenizer=model_class.tokenizer)

# peft can be in ["lora", "prefixtuning"]
trainer = CausalLMTrainer(train_dataset=train_dataset, 
                        validation_dataset=val_dataset, 
                        peft=None,
                        pretrained_model_or_path=model_class.get_model(),
                        tokenizer=model_class.get_tokenizer())
trainer.train()
\end{lstlisting}

\noindent The trainer handles data loading, tokenization, gradient accumulation, and checkpoint saving automatically. Users can customize training hyperparameters through the configuration dictionary or use sensible defaults.

\paragraph{Standardized Evaluation.}
CodeTF provides unified evaluation interfaces for popular benchmarks. The following example evaluates a CodeGen model on the HumanEval benchmark using the pass@k metric:

\begin{lstlisting}[language=Python]
from codetf.models import load_model_pipeline
from codetf.data_utility.human_eval_dataset import HumanEvalDataset
from codetf.performance.model_evaluator import ModelEvaluator

os.environ["HF_ALLOW_CODE_EVAL"] = "1"
os.environ["TOKENIZERS_PARALLELISM"] = "true"

model_class = load_model_pipeline(model_name="causal-lm", task="pretrained",
            model_type="codegen-350M-mono", is_eval=True,
            load_in_8bit=True, weight_sharding=False)

dataset = HumanEvalDataset(tokenizer=model_class.get_tokenizer())
prompt_token_ids, prompt_attention_masks, references= dataset.load()

problems = TensorDataset(prompt_token_ids, prompt_attention_masks)

evaluator = ModelEvaluator(model_class)
pass_at_k = evaluator.evaluate_pass_k(problems=problems, unit_tests=references, k=[1,10,100])
\end{lstlisting}

\noindent The evaluator handles code generation, test case execution, and metric computation. It supports parallel evaluation and provides detailed per-problem results in addition to aggregate metrics.

\section{Related Work}
\label{sec:related}

This section provides an overview of research on LLMs for code, discusses commercial code assistants, and compares CodeTF with related libraries and tools.

\paragraph{Large Language Models for Code.}
The application of large language models to code has grown rapidly, driven by their success on a wide range of tasks including code generation~\cite{codebert,codet5,elnaggar2021codetrans,to2023better}, code completion~\cite{codebert,codet5,peng2021could}, program repair~\cite{xia2022practical}, and code translation~\cite{roziere2020unsupervised,bui2019towards}. Building on the success of BERT~\citep{bert} and GPT~\citep{gpt} in natural language processing, researchers have developed specialized models that treat code as a sequence of tokens and leverage pretraining strategies such as masked language modeling, span corruption, and causal language modeling~\cite{codebert,codet5,guo2020graphcodebert,ahmad2021unified,elnaggar2021codetrans,peng2021could,kanade2020learning,chakraborty2022natgen,ahmed2022multilingual,niu2022spt, bui2022detect,bui2021self,bui2021infercode,huynh2022hierarchynet}.

\begin{table}[t]
	\centering
	\label{tab:comparison}
	\caption{Comparison of features between CodeTF and HuggingFace Transformers (HF-T).
    Note that we compare these libraries by features related to the code domain, highlighting functionalities that HF-T may not specifically support.
 }
	\begin{tabularx}{\textwidth}{>{\raggedright\arraybackslash}Xcc}
		\toprule
		Feature & CodeTF (Ours) & HF-T \\
		\midrule
		Unified Model and Dataset Interface & \cmark & \cmark \\
		Unified Parameter-Efficient Fine-Tuning for Code Intelligence Tasks & \cmark & \cmark \\
		Unified Code Utility Interface for Multiple Programming Languages & \cmark & \xmark \\
		Unified Metric Interface to Evaluate Code Intelligence Benchmarks & \cmark & \xmark \\
		Unified Data Loader Interface to Process Code Intelligence Benchmarks & \cmark & \xmark \\
		Modular Library Design & \cmark & \cmark \\
		Pretrained Model Checkpoints & \cmark & \cmark \\
		Task-specific Fine-tuned Model Checkpoints & \cmark & \cmark \\
		\bottomrule
	\end{tabularx}
\end{table}

Code LLMs can be categorized into three primary architectures: encoder-only models~\citep{codebert,graphcodebert,codemvp}, decoder-only models~\citep{codexglue,codex,incoder,codegen}, and encoder-decoder models~\citep{plbart,codet5,sptcode,natgen}. Encoder-only models excel at understanding tasks such as code retrieval~\citep{csn} and clone detection, while decoder-only models are well-suited for generation tasks like program synthesis~\citep{codex,apps}. Encoder-decoder models~\citep{codet5,plbart} offer versatility for both understanding and generation but may not always outperform specialized architectures on individual tasks. CodeTF supports all three architecture families, enabling users to select the most appropriate model for their specific use case.

\paragraph{Commercial Code Assistants.}
The commercial impact of Code LLMs has been transformative. GitHub Copilot, powered by OpenAI's Codex model, was launched in 2021 and has been adopted by millions of developers, demonstrating significant productivity improvements in real-world software development. Amazon CodeWhisperer and Google's code-assist tools offer similar capabilities for their respective cloud ecosystems. Tabnine provides IDE-integrated code completion using both cloud and local models, while Replit's Ghostwriter focuses on the educational and hobby developer market.

While these commercial tools have proven the value of Code LLMs in practice, they present limitations for research and customization: (1) they are closed-source, preventing examination or modification of their underlying models; (2) they do not support fine-tuning on custom datasets or domains; (3) they lack transparency in evaluation, making it difficult to compare with published research results. CodeTF addresses these limitations by providing an open-source framework with full access to model weights, training pipelines, and evaluation code.

\paragraph{Open-Source Libraries for Code Intelligence.}
Several open-source libraries have been developed to support Code LLM research and development. NaturalCC~\cite{wan2022naturalcc} is a platform designed to facilitate NLP-based code analysis research, offering utilities for training and reproduction. However, its usability is limited by complex dependencies and configuration requirements.

HuggingFace Transformers~\cite{HuggingfaceTransformers} is the most widely-used library for working with pretrained language models, offering user-friendly interfaces across multiple domains (vision, language, code, time series). While HuggingFace provides excellent general infrastructure, it lacks code-specific features such as AST parsing, code-aware evaluation metrics, and specialized preprocessing utilities.

Other repositories such as CodeT5~\cite{codet5}, CodeGeeX~\cite{zheng2023codegeex}, CodeBERT~\cite{codebert}, and CodeXGLUE~\cite{codexglue} provide model implementations and benchmark code, but these are typically model-specific rather than unified libraries. The BigCode project has made valuable contributions through open-source models (StarCoder, SantaCoder) and datasets (The Stack), but does not provide a unified training and evaluation framework.

\paragraph{Positioning of CodeTF.}
Table~\ref{tab:comparison} summarizes the comparison between CodeTF and HuggingFace Transformers for code intelligence tasks. While HuggingFace Transformers provides comprehensive general-purpose infrastructure, CodeTF offers specialized features for the code domain: a code utility interface with built-in AST parsing for 15+ languages, code attribute extraction, and language-specific preprocessing capabilities not available in general NLP libraries; standardized implementations of code-specific evaluation metrics (pass@k, CodeBLEU, Edit Similarity) with benchmark-specific evaluation scripts; and pre-built data loaders for HumanEval, MBPP, APPS, CodeXGLUE, and The Vault with consistent interfaces. CodeTF is designed as a complementary layer built upon HuggingFace Transformers rather than a replacement, leveraging its robust model infrastructure while adding code-specific capabilities that simplify code intelligence research and development.

\section{Future Plans and Improvements}
\label{sec:future}

We continue to actively improve CodeTF as a one-stop open-source library for Code LLMs and code intelligence tasks. We have several plans to expand its capabilities and support more advanced use cases and improve model reproducibility. Some key features we aim to incorporate in the future include:

\begin{itemize}[leftmargin=*]
	\item Implementing 4-bit quantization as part of the pretrained and fine-tuned models, enabling even large models such as InstructCodeT5+ \cite{wang2023codet5+} to run efficiently on commercial laptops or workstations.
	\item Conducting comprehensive evaluations of well-known code intelligence tasks on established benchmarks (CodeXGLUE, MBPP, HumanEval, and APPS). Due to rapid advancements in the field, there is a lack of reproducibility in the performance of state-of-the-art models, making it challenging for the research community to adapt and foster collaboration.
	\item Enhancing the Code Utility module by adding support for other programming languages, such as Go, Rust, C\#, and more. We also plan to include utilities for extracting additional useful features from code, such as call graphs, control flow, data flow, and others.
	\item Integrating a broader selection of recent state-of-the-art pretrained language models of code into CodeTF, further solidifying our library as a comprehensive resource in the field.
\end{itemize}

\section{Conclusion}
\label{sec:conclusion}

We have presented CodeTF, a comprehensive open-source library for code intelligence that addresses the fragmented landscape of Code LLM development and deployment. By providing unified interfaces for model loading, training, evaluation, and code-specific utilities, CodeTF significantly reduces the engineering effort required to work with state-of-the-art Code LLMs.

The key technical contributions of CodeTF include a unified model interface supporting 9+ pretrained models across three architectures (encoder-only, decoder-only, encoder-decoder) with seamless quantization for efficient deployment; comprehensive code utilities including AST parsing for 15+ programming languages, enabling code attribute extraction and language-aware preprocessing without external dependencies; standardized evaluation infrastructure with built-in support for popular benchmarks (HumanEval, MBPP, APPS, CodeXGLUE) and consistent metric implementations (pass@k, CodeBLEU, Edit Similarity); and parameter-efficient fine-tuning integration through HuggingFace PEFT, enabling adaptation of large models on consumer hardware.

CodeTF is designed to serve both the research community exploring new Code LLM architectures and practitioners deploying models in production systems. We believe that providing standardized, reproducible infrastructure will accelerate progress in code intelligence research by enabling fair comparisons across models and reducing the barrier to entry for new researchers.

We invite the community to contribute to CodeTF's continued development. Contributions may include adding support for new models and benchmarks, implementing additional programming language parsers, improving documentation, and sharing evaluation results. The library is available at \url{https://github.com/salesforce/CodeTF} under an open-source license.

As Code LLMs continue to advance rapidly, we are committed to maintaining CodeTF as a comprehensive and up-to-date resource. Future development will focus on integrating emerging models, expanding language support, and adding new evaluation capabilities as the field evolves.

\section{Broader Impact and Responsible Use}
\label{sec:broader_impact}
While models within CodeTF show immense potential in various code-related tasks, they do not provide absolute guarantees regarding their code intelligence capabilities. The datasets and pretrained models used in CodeTF may carry biases that could result in misinterpretations, incorrect results, or undesired behaviors. These biases can take multiple forms:

\begin{enumerate}[leftmargin=*]
	\item \textbf{Language Bias}: The model might prefer certain programming languages over others based on the frequency of the languages in the training data. For instance, if the model is trained mostly on Python code, it might struggle to generate accurate and idiomatic Java or JavaScript code. 
	\item \textbf{Application-specific Bias}: This occurs when a model trained for a particular application or domain is used in a different application. For example, a model trained on web development code may perform poorly when tasked with generating embedded system code.
	\item \textbf{Library and Framework Bias}: This refers to the inherent inclination of a model towards using specific libraries or frameworks due to the frequency of their presence in the training dataset. For example, if the model was predominantly trained on data using Python's Pandas for data manipulation, it may be more inclined to use Pandas even in situations where other libraries like NumPy or native Python constructs could be more efficient or appropriate.
	\item \textbf{Language Version Bias}: Software languages evolve, with new versions (e.g., Python 2 to Python 3) introducing changes, depreciations, and novel features. If the training dataset is not updated regularly to reflect these changes, the model could generate code using outdated or deprecated conventions of a language.
	\item \textbf{Coding Style Bias}: Coding style can vary significantly between individual coders, teams, or communities. If the model is trained predominantly on a dataset reflecting a specific style, it may generate code that is in accordance with that style, which may not be the optimal or preferred way for the specific use-case at hand.
	\item \textbf{Solution Bias}: There can often be more than one valid solution to a coding problem. The model might be biased towards the solutions it was exposed to during training and might fail to generate other potentially more efficient or elegant solutions.
\end{enumerate}

In addition to these potential biases, there are several other crucial considerations:

\begin{enumerate}[leftmargin=*]
	\item \textbf{Sustainability}: Energy efficiency is a significant concern in AI, especially with large-scale models. Optimized models generating more efficient code could reduce the computational resources required to execute such code, thereby reducing energy consumption. Ongoing research into more energy-efficient AI training methods can also decrease the energy footprint of AI itself.
	\item \textbf{Inclusive language}: Coding language needs to be inclusive as the field becomes increasingly diverse. Non-inclusive terms may discourage and offend many developers. Future work should focus on creating tools to identify non-inclusive language in code and recommend suitable alternatives.
	\item \textbf{Job loss and automation}: While AI carries the potential to automate certain tasks, it is essential to view it as a tool that augments rather than replaces human efforts. Developer tools are usually designed to handle repetitive tasks, freeing developers to focus on complex issues. However, it's crucial to ensure developers do not become overly reliant on these tools and can still code effectively on their own.
	\item \textbf{Human control and autonomy}: Maintaining human control and oversight is crucial, especially in critical areas like code generation. Techniques like explainability and interpretability in AI, along with rigorous testing, ensure AI systems remain under human control and behave as expected. The goal should be to create AI systems that enhance human capabilities and work collaboratively with humans, rather than replacing them.
\end{enumerate}

Users of CodeTF must scrutinize the pretrained models and the general system before their adoption in practical applications. We are committed to refining the library by identifying and addressing potential biases and inappropriate behaviors continually.

We encourage researchers, software engineers, and AI practitioners to use the library responsibly for applications that enhance software quality and developer productivity. However, CodeTF should not be used to develop code intelligence models that could lead to unethical capabilities, such as unauthorized code manipulation, privacy breaches, or the propagation of insecure coding practices.

As AI becomes more integrated into software development, it is essential to address these ethical and practical considerations. CodeTF is committed to supporting responsible AI practices and mitigating potential biases and inappropriate behaviors moving forward.

\bibliographystyle{unsrt}  
\bibliography{custom}  

\end{document}